\documentclass[preprint,12pt]{elsarticle}

\usepackage[latin1]{inputenc} 
\usepackage[english]{babel} 
\usepackage[T1]{fontenc} 
\usepackage{charter} 
\usepackage{graphicx} 
\usepackage{textcomp} 
\usepackage{wrapfig} 
\usepackage{amsmath} 
\usepackage{amssymb} 
\usepackage[version=3, arrows=pgf]{mhchem} 
\usepackage{layout} 
\usepackage[a4paper, bottom=3cm]{geometry} 
\usepackage[font=scriptsize]{caption} 
\usepackage[font=footnotesize]{subcaption} 
\usepackage{setspace} 
\usepackage{gensymb} 
\usepackage{titlesec} 
\usepackage{float} 
\usepackage{pdfpages} 
\usepackage{fancyhdr} 
\usepackage{hyperref} 
\usepackage{tabularx} 
\usepackage{comment} 
\usepackage{multirow} 
\usepackage[colorinlistoftodos]{todonotes}

\providecommand{\e}[1]{\ensuremath{\times 10^{#1}}}

\let\oldtabular\tabular
\renewcommand{\tabular}{\footnotesize\oldtabular}

\usepackage{graphicx}
\usepackage{amssymb}

\journal{Journal of the European Ceramic Society}

\begin{document}

\begin{frontmatter}


\title{Fabrication of ice-templated tubes by rotational freezing: microstructure, strength, and permeability}

\author[CREE]{Jordi Seuba}
\author[CREE]{Jerome Leloup}
\author[CREE]{Stephane Richaud}
\author[CREE]{Sylvain Deville}
\author[IEM]{Christian Guizard}
\author[CREE]{Adam J. Stevenson}

\address[CREE]{Ceramics Synthesis and Functionnalization Lab, UMR 3080 CNRS/Saint-Gobain, F-84306 Cavaillon, France.}
\address[IEM]{Institut Europeen des Membranes, Universite de Montpellier 2, Place Eugene Bataillon, 34095 Montpellier Cedex 5, France.}

\begin{abstract}

We demonstrate a facile and scalable technique, rotational freezing, to produce porous tubular ceramic supports with radially aligned porosity. The method is based on a conventional ice-templating process in a rotatory mold and demonstrated here with yttria-stabilized zirconia (YSZ). We investigated the effects of solid loading, freezing temperature, and volume of the slurry on the microstructure, strength (o-ring test and four-point bending), and air permeability. The results show that pore volume and pore size can be controlled by the solid loading and freezing temperature respectively, and overall tube thickness can be adjusted by the volume of slurry initially poured into the mold. Decreasing pore size and pore volume increases the mechanical properties but decreases the air permeability. These tubes could be particularly interesting as tubular membrane supports such as oxygen transport membranes.

\end{abstract}

\begin{keyword}
Ice-templating \sep Ceramics \sep Tubular membranes \sep Fracture \sep Air flow

\end{keyword}

\end{frontmatter}


\section{Introduction}

Porous ceramic membranes are increasingly used in technologically important applications such as energy production (solid oxide fuel cells, oxygen and proton transport membranes) \cite{Pan2005} \cite{Guo2013} \cite{Sunarso2008} \cite{Weirich2012}, solid-liquid filtration \cite{Barg2011} \cite{Ambrosi2014}, and gas separation \cite{Zivkovic2004}. These membranes usually have an asymmetric configuration where a membrane layer is deposited on top of a porous substrate. The membrane layer provides the selectivity required for the process, while the porous support provides the mechanical stability. Besides strength, porous supports must also exhibit a high permeability to facilitate the evacuation of the fluid and thus minimize detrimental effects such as concentration polarization or exaggerated pressure gradients that compromise the overall performance of the membrane \cite{Baumann2013}.   

Typically, tubular porous substrates are prepared by extrusion \cite{Zhou2011} \cite{Dong2007}, slip casting \cite{choi2012oxygen}, centrifugal casting \cite{Kim2002}, or isostatic pressing \cite{dong2009reaction}. However, these techniques rely on increasing the total pore volume and pore size to achieve the required permeability at the expense of mechanical properties. Therefore, it is necessary to explore other strategies such as an optimized pore shape or reduced tortuosity to simultaneously improve permeability and strength.  

Nowadays, the method to obtain unidirectional macroporous materials most conventionally used is extrusion. However due to technical limitations is not possible to produce materials with a high pore volume content (>50\%) and reduce the pore size below 100 $\mu m$ \citep{Isobe2007}. On the other hand, direct replication of wood has successfully been demonstrated as a feasible technique, although the large number of steps and the microstructural variability among samples remain as important drawbacks to overcome. Alternatively, in the last decade ice-templating (or freeze-casting) has emerged as a popular shaping route to produce macroporous materials with a tailored pore morphology and aligned porosity. The technique is based on the freezing of a colloidal suspension and the segregation of ceramic particles between the crystals. Afterwards, the solvent crystals are removed by sublimation and the green body is sintered to consolidate the structure. The final porosity is thus a direct replica of the solvent crystals \cite{Deville2008}.  

The mechanical and gas flow properties of ice-templated monoliths have been extensively evaluated \cite{Deville2015} \cite{seuba2016a} \cite{Porter2014} \cite{Chen2007a} \cite{Han2010} \cite{Xue2012} \cite{seuba2016b} \cite{Fukushima2010} \cite{Pekor2010}. However the literature on ice-templated tubes is scarce. Moon et al. \cite{Moon2003} used an external metallic mold with a central PTFE rod to induce the thermal gradient required to align the porosity. They studied the effect of solids loading and freezing temperature on the microstructure and the feasibility of depositing an external dense layer. Liu et al. \cite{Liu2013f} used a similar set-up to evaluate the impact of solids loading, pore size, and particle size on compressive strength, nitrogen flux, and water flux. However, the difficulties demolding the tubes after solidification is drawback of this set-up that limits further industrial development.  Moon et al. \cite{Moon2012} overcomes this issue combining ice-templating with co-extrusion, however the resultant porosity is not radially aligned.

In this work, we propose a novel method based on a rotational freezing to produce ice-templated tubes. We investigate the impact of pore volume, size, and tube thickness on mechanical properties using an o-ring test and four-point bending. Furthermore, we studied the effect of the same morphological parameters on permeability to assess the applicability of these materials as a membrane supports. 

\section{Experimental procedure}
\subsection{Sample preparation}

Ceramic suspensions were prepared by mixing distilled water with 0,75 wt\% of dispersant (Prox B03, Synthron, Levallois-Paris, France), 3 wt\% of organic binder PVA (PVA2810, Wacker, Burghausen, Germany), and 3 mol\% yttria-stabilized zirconia (TZ-3YS, Tosoh, Tokyo, Japan). All the percentages are referred to the total solids loading of the slurry. After a first mixing step with a magnetic stirrer, the slurry was ball milled at 800 rpm for a minimum of 18 h to ensure a good homogenization. Afterwards, it was deaired during 10 minutes and poured in a tubular cooper mold.

The set-up used to ice-template tubes consists of a cryothermostat (Model CC 905,
Hubert, Offenburg, Germany), freezing liquid (SilOil M90.055.03, Hubert, Offenburg, Germany), metallic container, tube-shape mold, and a rotating system. The tube filled with slurry was attached to the rotational clamp and the speed set at 70 rpm. The silicon oil was pumped in the container with a preset temperature (-80$^{\circ}$C or -30$^{\circ}$C) until it achieved a steady flow. The system was left under rotational freezing for a minimum of 5 minutes. The next steps were similar to the regular ice-templating process and detailed elsewhere \cite{seuba2016a}: demoulding, freeze-drying (Free Zone 2.5 Plus, Labconco, Kansas City, Missouri, USA), and sintering. Tubular specimens were sintered in air at 1400$^{\circ}$C for 3 h with a heating and cooling rate of 5 $^{\circ}$C/min. A previous step at 500$^{\circ}$C for 5 h with the same heating rate was used to ensure a proper organic burn-out.

The process was repeated with variations of the solid loading of the initial slurry (50 wt.\%, 55 wt. \%, and 65 wt.\%), temperature of the freezing liquid (-30 $^{\circ}$C and -80$^{\circ}$C), and volume of slurry initially poured into the mold (16, 18, and 20 ml). The final dimensions of the tubes in all tested conditions were: 150 mm length (\textit{L}) and 10 mm of external diameter ($D_{ext}$).

\subsection{Sample characterization}

Micrographs of the ice-templated tubes were taken with a scanning electron microscope (Nova NanoSEM 230, FEI, Hillsboro, USA) at 10-15 kV. Total pore volume P(\%) was calculated based on the relative density measured geometrically with respect to that of fully dense TZ-3YS ($\rho_{ysz} = 5.8$ $gcm^{-3}$). P(\%) measurements were performed in a minimum of five rings per tube, and two tubes per condition. The average pore size was measured by mercury intrusion porosimetry (AutoPore IV 9500, Micromeritics, Norcross, USA) with an applied pressure up to 0,31 bar.

Mechanical properties of ice-templated tubes were characterized by two different tests: 1) four-point bending on half tubes, and 2) o-ring test. All the experiments were conducted with a universal testing machine (Shimadzu model AGSX, Kyoto, Japan), equipped with a 10 kN load cell at a crosshead speed of 0.2 mm/min.

\begin{enumerate}
\item Samples were cut with a slow-speed saw first in the radial direction and then in the longitudinal direction with a final length (\textit{L}) $\geq$ 55 mm, and external diameter ($D_{ext}$) = 10 mm. The spacing between the outer supports was adjusted to $S_{1}$ = 40 mm whereas the span between the inner supports was $S_{2}$ = 20 mm. A minimum of three samples per condition were tested.

The maximum stress ($\sigma_{max}$) at the middle point between the two inner supports was calculated based on the approximation proposed by Kwok et al. \cite{Kwok2014}. Tests were considered invalid when the crack propagated outside the inner supports. When the test crack pattern resulted in a successful test, the maximum stress was calculated by:

\begin{equation}
\label{eq:Bending1}
\sigma_{max} = \frac{Pay_{c}}{I}
\end{equation}

where \textit{P} is the maximum load, \textit{a} is the distance between the load and the support, $y_{c}$ is the distance between the middle point and the centroid of the specimen cross section along the radial direction, and \textit{I} is the second moment of inertia of the cross section. For a semi-circular annulus, $y_{c}$ and \textit{I} are expressed as: 

\begin{equation}
\label{eq:Bending3}
y_{c} = R - \frac{4}{3\pi}\frac{R^{3}-{R_{i}}^{3}}{R^{2}-{R_{i}}^{2}} 
\end{equation}

and

\begin{equation}
\label{eq:Bending2}
I = \frac{\pi}{8}(R^{4}-{R_{i}}^{4})-\frac{8}{9\pi}\frac{(R^{3}-{R_{i}}^{3})^{2}}{R^{2}-{R_{i}}^{2}}
\end{equation}

where \textit{R} and \textit{$R_{int}$} are the external and internal radius respectively.

\item Eight tubes were sliced in the radial direction with a slow-speed saw leaving specimens  with an external diameter ($D_{ext}$) = 10 mm, height (\textit{h}) = 4 mm, and ring thickness (\textit{t}) $\approx$ 2 mm. Tests were performed with a piece of alumina supporting the samples to avoid the rotation and distribute evenly the load through the surface.

The maximum stress ($\sigma_{max}$) is located on the inner diameter of the ring and was determined by:

\begin{equation}
\label{eq:O-Ring}
\sigma_{max} = 1,8 \frac{Pr_{a}}{ht} \left(1 + \frac{t}{3r_{a}}\right)
\end{equation}

where \textit{P} is the maximum load, \textit{t} the tube thickness, \textit{h} the width of the ring, and \textit{$r_{a}$} the average radius $(r_{ext}-r_{int})/2$.

\end{enumerate}
Gas permeability was evaluated using custom built equipment to measure the air pressure drop across the tubes. A preliminary test was performed to assess the tightness of the system. Synthetic air was passed through the tubular samples which were held by a silicon ring. The specimens exhibited one closed side, 120 mm length (\textit{L}), and 10 mm external diameter ($D_{ext}$). The air flow studied ranged between 0-25000 ml/min (0-0.25 m/s) and was controlled by an electrovalve. Two sensors were placed before and after the tube to record the inlet ($P_{i}$) and outlet pressure ($P_{o}$).

\section{Results and discussion}
\subsection{Microstructure control}

\begin{figure}[H]
\centering\includegraphics[width=0.6\textwidth]{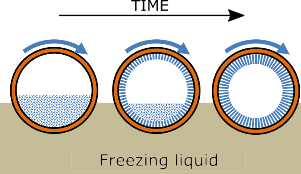}
\caption{Schematic of the rotational freezing process.}
\label{fig:Rotation}
\end{figure}

Fig. \ref{fig:Rotation} schematically depicts the process that occurs during the rotational freezing. First, the tube rotation spreads the slurry evenly across the inner surface of the mold. Then, the freezing oil is pumped into the container and the slurry in contact with the mold solidifies almost instantly. The freezing rate at this initial stage is very high due to the high heat transfer between the copper mold and the slurry. Afterwards, the velocity of the solidification front undergoes an abrupt decrease, until it reaches a steady rate. 

\begin{figure}[H]
\centering\includegraphics[width=0.95\textwidth]{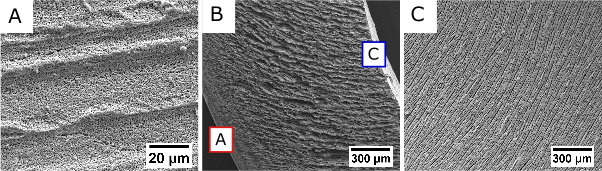}
\caption{SEM micrographs at different locations of an ice-templated tube with a 65 wt.\% solid loading, frozen at -30$^{\circ}$C, and 20 ml. A) Top view of the external layer B) Cross section of the tube and C) Top view of the inner surface.}
\label{fig:Micro}
\end{figure}

The ice-templated tubes exhibited two different pore morphologies, random (area in contact with the copper mold) and lamellar porosity, similarly to that reported in unidirectional frozen ice-templated monoliths \cite{Deville2010b}. The external layer with random porosity is generated when the velocity of the solidification front is too high for the ice crystals to grow with a preferential orientation and thus they are not able to rearrange the ceramic particles (Fig. \ref{fig:Micro}-A). When steady state is reached, the ice crystals grow continuously along the thermal gradient (i.e. in this case the radial direction) and the pore morphology becomes lamellar (Fig. \ref{fig:Micro}-B). Due to the constant rotation of the system, the remaining liquid remains in contact with the frozen part and provides the slurry required to continue with the ice-templating process until the tube is entirely solidified.

\begin{figure}[H]
\centering\includegraphics[width=0.6\textwidth]{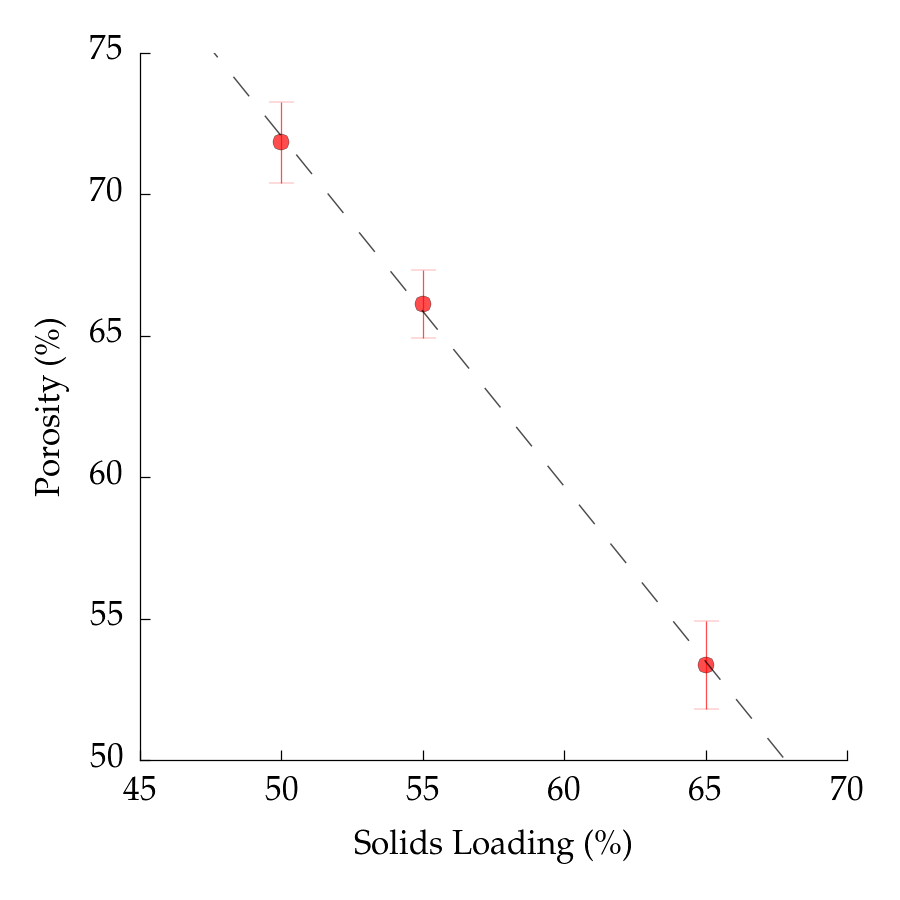}
\caption{Variation of porosity as a function of the solid loading of the slurry. Dashed line represents the the relationship between solid loading and porosity obtained in a previous work for ice-templated YSZ monoliths \cite{seuba2016a}.}
\label{fig:Porosity}
\end{figure}

Fig. \ref{fig:Porosity} shows the effect of solid loading on total pore volume. Increasing the solid loading causes a variation in the ceramic particles-water ratio and therefore an increase in the relative density. The total pore volume of the ice-templated tubes ranged between 53\% and 73\% for a variation in solid loading from 65\% to 50\%. The dashed line in Fig. \ref{fig:Porosity} represents the relationship between solid loading and porosity obtained in a previous work for ice-templated YSZ monoliths \cite{seuba2016a}. The good match between the monolithic samples and the tubular samples shows that total pore volume is independent of the mold used and mainly determined by solid loading. Since the pore volume measurements were taken at different locations along the tube, it also indicates that the pore volume is homogeneously distributed along the length of the tube.

During ice-templating, pore size can be controlled by the freezing temperature and solids loading. Table \ref{tab:Morpho} shows how, for a given solids loading, samples frozen at -80$^{\circ}$C systematically exhibited a smaller pore size compared to their counterparts frozen at -30$^{\circ}$. At lower freezing temperatures, the temperature gradient and the velocity of the freezing front increase,  which result in smaller ice crystals and creates thus a finer microstructure \cite{Farhangdoust2014}. Furthermore, solids loading also affects the pore size. This phenomena is caused by the thicker walls created when the solid loading increases hindering the ice crystal growth.

\begin{table}[H]
\centering
\resizebox{0.9\textwidth}{!}{\begin{tabular}{c c c c c}
	\hline
	\textbf{Solids loading (wt.\%)} & \textbf{Porosity (\%)} & \textbf{Freezing Temperature ($^{\circ}$C)} & \textbf{Mean $d_{p}$ ($\mu m$)} & \textbf{$\sigma_{ORing}$ (MPa)} 
    \\ \hline \hline
	65 & 53.96 $\pm$ 2.54 & -80 & 3.3 & 20.81 $\pm$ 6.6\\
	55 & 65.70 $\pm$ 0.83 & -80 & 6.8 & 14.01 $\pm$ 3.90 \\
	50 & 70.89 $\pm$ 1.55 & -80 & 7.3 & 8.64 $\pm$ 3.36\\
	\hline
	65 & 52.64 $\pm$ 1.85 & -30 & 6.2 & 28.31 $\pm$ 4.86\\
	55 & 67.64 $\pm$ 1.18 & -30 & 10.6 & 15.98 $\pm$ 1.39 \\
	50 & 72.50 $\pm$ 0.96 & -30 & 14.6 &  7.63 $\pm$ 1.31\\	
	\hline
\end{tabular}}
\caption{Summary of the effect of solids loading and freezing temperature on porosity, mean pore size ($d_{p}$), and radial crushing strength ($\sigma_{ORing}$). $d_{p}$ was obtained by mercury intrusion porosimetry.}
\label{tab:Morpho} 
\end{table} 

\begin{figure}[H]
	\centering
	\includegraphics[width=0.6\textwidth]{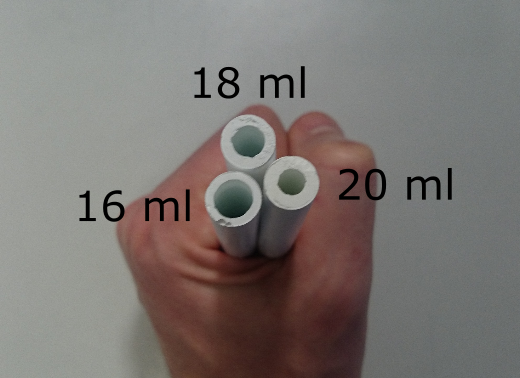}
	\caption{Detail of the thickness variation by changing the initial amount of slurry.}
	\label{fig:Thickness}
\end{figure}

Fig. \ref{fig:Thickness} shows a top view of three ice-templated tubes with a different amount of slurry initially poured into the mold (20, 18, and 16 ml). Decreasing the volume of slurry reduces  the overall thickness of the tube because essentially there is less material to be ice-templated. The overall thickness of the tubes after sintering was 2.0, 1.8, and 1.6 mm for slurry volumes of 20, 18, and 16 ml.

\subsection{Mechanical properties}
\subsubsection{O-Ring test}

\begin{figure}[H]
	\centering
	\includegraphics[width=0.9\textwidth]{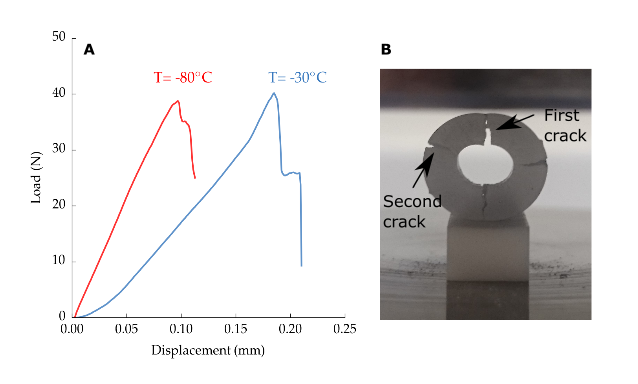}
	\caption{A) Representative load-displacement curves of two rings obtained from tubes with a 65\% solids loading and frozen at -80$^{\circ}$C (red) and -30$^{\circ}$C (blue). B) Detail of the characteristic fracture in a O-ring test with the two types of cracks highlighted.}
	\label{fig:Fracture_Ring_panel}
\end{figure}

Fig. \ref{fig:Fracture_Ring_panel}-A shows two representative load-displacement curves for samples frozen at -80$^{\circ}$C and -30$^{\circ}$C. In both cases, the fracture behavior is described by an initial linear stage followed by two consecutive load drops. When the ring is subjected to a diametrical compression i.e. like in the o-ring test, the maximum tensile stresses are located on the intersection between the inner diameter and the loading plane \cite{Hudson1969}. This load corresponds to the first peak in the load-displacement curve and was used to describe the strength of the rings. The second peak corresponds to the fracture of the two remaining "\textit{half rings}", originated by the tension developed on the outer periphery. Both types of cracks can be observed in Fig. \ref{fig:Fracture_Ring_panel}-B. 

\begin{figure}[H]
	\centering
	\includegraphics[width=0.6\textwidth]{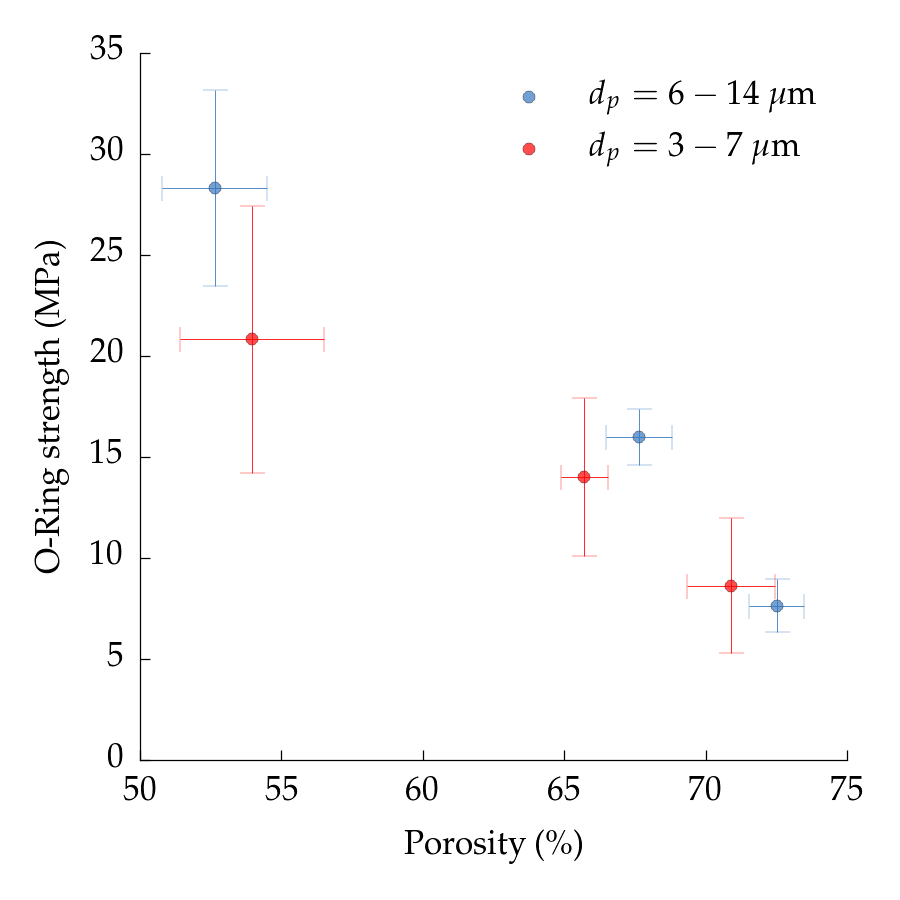}
	\caption{Effect of porosity and freezing temperature on radial crushing strength obtained by O-ring test. Data summarized in Table \ref{tab:Morpho}.}
	\label{fig:Oring}
\end{figure}

Fig. \ref{fig:Oring} and Table \ref{tab:Morpho} show the effect of pore volume on strength for samples frozen at -80$^{\circ}$C and -30$^{\circ}$C. As expected, rings with higher pore volume exhibited a lower strength. Moreover, samples frozen at a lower temperature (-80$^{\circ}$C) have a smaller pore size and therefore should exhibit an increase in strength \cite{Brezny1991} \cite{seuba2016a}. However, the trend obtained in this study seemingly contradicts this observation. The explanation for this behavior could rely on the large scattering of the strength data that makes it difficult to conclude whether this effect is seen here or not.

\begin{figure}[H]
	\centering
	\includegraphics[width=0.95\textwidth]{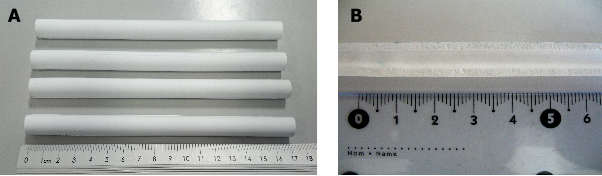}
	\caption{Examples of tubes made by rotational freezing. A) Overview B) Cross section of a tube.}
	\label{fig:tubes_pics}
\end{figure}

There are two contributions to the scattering of strength values: the probabilistic nature of the strength in ceramics \cite{Danzer1992}, and small, local variations of the tube thickness, as observed in Fig. \ref{fig:tubes_pics}-B. Such geometrical imperfections induce an asymmetry in the stress field that have a deleterious effect on the strength \cite{Kwok2014b}, and in this case may hide the pore size effect.

\subsubsection{Four-point bending}

\begin{figure}[H]
	\centering
	\includegraphics[width=\textwidth]{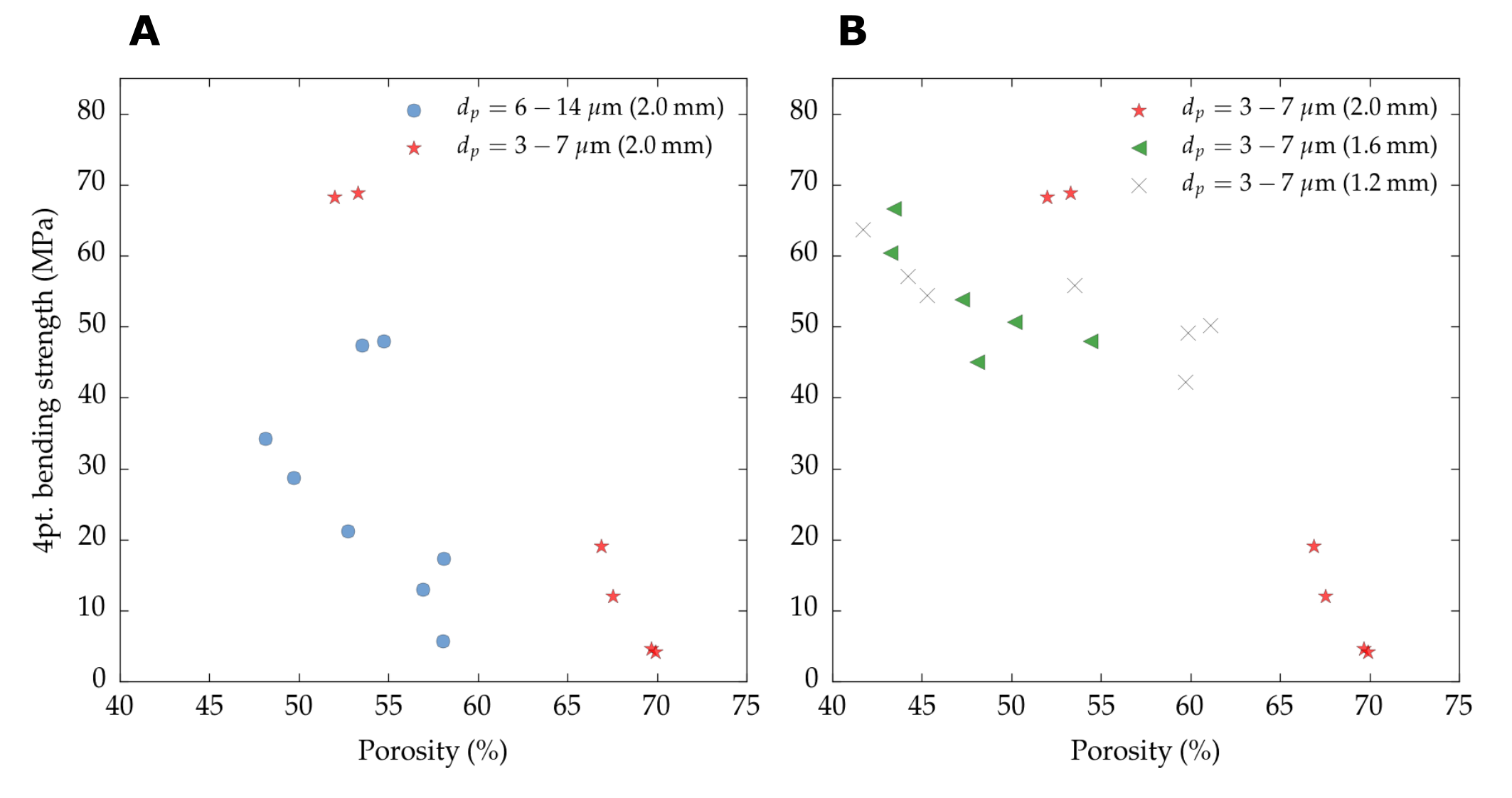}
	\caption{Effect of freezing temperature A), and volume of slurry (which determines the tube thickness) B) on flexural strength obtained by four-point bending.}
	\label{fig:Bending}
\end{figure}

We modified the solid loading, freezing temperature, and volume of the slurry initially poured into the mold to obtain ice-templated tubes with different pore volume, pore size, and thickness respectively. Fig. \ref{fig:Bending} shows the effect of each parameter on the flexural strength.   

In all conditions, when the solid loading increases the total pore volume of the sample decreases, causing an increase in strength. Unlike in the diametrical compression, in four-point bending we did observe an increase of strength with pore size reduction (Fig. \ref{fig:Bending}-A). We attributed this effect to the increasing amount of ceramic struts between the adjacent walls observed at lower freezing temperatures \cite{Munch2009a} that reinforce the structure. When the half ice-templated tube is bent, the external midpoint is subjected to tensile forces. Since the porosity is perpendicular to the tensile forces, the strength of the tube is mainly determined by the resistance of the structure to opening the pores.

On the other hand, it is more difficult to discern the effect of tube thickness on flexural strength (Fig. \ref{fig:Bending}-Bb). As discussed above, the strength data are too scattered to drawn any conclusion due to local imperfections of the tubes geometry.

\subsection{Permeability}

\begin{figure}[H]
	\centering
	\includegraphics[width=\textwidth]{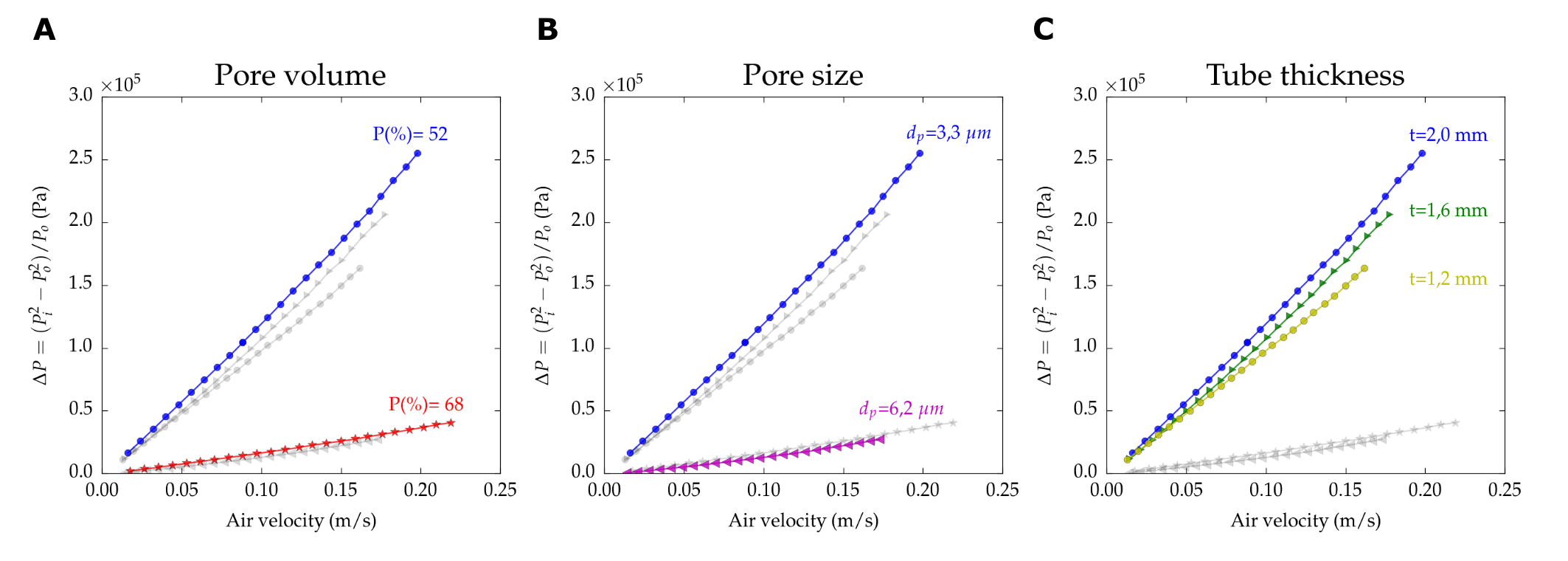}
	\caption{Effect of experimental conditions on pressure drop. A) 65\% and 50\% solid loading, B) -80$^{\circ}$C and -30$^{\circ}$C, and C) 20, 18, and 16 ml poured initially in the mold. Values specified in Table \ref{tab:StrucVSPermea}.}
	\label{fig:PressureDrop_Panel}
\end{figure}

Tubes obtained with different solids loading, freezing temperature, and volume of slurry poured into the mold were tested to evaluate the independent impact of pore volume, pore size, and tube thickness respectively on permeability, Fig. \ref{fig:PressureDrop_Panel}. 

In all conditions, pressure drop ($\Delta P$) exhibits a linear dependence with air velocity ($\nu_{s}$). A similar behavior was observed in a previous work \cite{seuba2016b} with ice-templated YSZ monoliths in the same pore volume and pore size range. Therefore, permeability can be calculated from their slopes according to the Darcy's law (Eq. \ref{eq:Darcy}) instead of the Forchheimer equation \cite{Innocentini1999}.
  
\begin{equation}
\label{eq:Darcy}
\frac{\Delta P}{L} = \frac{{P_{i}}^2 - {P_{o}}^2}{2P_{o} L} = \frac{\mu}{k_{1}}\nu_{s}
\end{equation}

\textit{L} is defined as the thickness of the tube, $\mu$ the viscosity of the fluid (in this case air at 25$^{\circ}$C), $\Delta P$ is the pressure drop, $\nu_{s}$ the fluid velocity (air flow divided by the internal area of the tube), and $k_{1}$ the Darcian permeability.

Permeability values ($k_{1}$), experimental conditions, and the most relevant pore descriptors are specified in Table \ref{tab:StrucVSPermea}. When the pore volume increases due to the solid loading reduction, there is less resistance to the airflow and permeability increases accordingly, Fig. \ref{fig:PressureDrop_Panel}-A. Additionally, increasing pore volume increases the mean pore size, enhancing further this effect. 

Fig. \ref{fig:PressureDrop_Panel}-B shows how the specimen frozen at -80$^{\circ}$C exhibits a smaller pore size (3.3 $\mu m$) and lower permeability (2.80\e{-14} $m^{2}$). When the tubes are frozen using a constant solid loading and lower temperature, the total pore volume remains invariant and the pore size reduces, affecting the permeability. Interestingly, the sample frozen at higher temperature (-30$^{\circ}$C), pore size of 6.2 $\mu m$, and 52\% porosity in Fig. \ref{fig:PressureDrop_Panel}-B has a comparable permeability (and hence pressure drop) to the sample frozen at -80$^{\circ}$C, pore size 7.3 $\mu m$, and porosity 68\% in Fig. \ref{fig:PressureDrop_Panel}-A. This result emphasizes the importance of pore size over pore volume as the critical parameter to tailor permeability. 

This observation is in agreement with the Hagen-Poiseuille equation that predicts a gas flow dependence (i.e. in a viscous flow) proportional to the pore volume compared with the square exhibited by pore size \cite{Ohji2012}. Finally, permeability can also be controlled by the volume of slurry initially poured into the mold (Fig. \ref{fig:PressureDrop_Panel}-C). When the volume of slurry decreases, the overall thickness of the tube decreases as well and the tube becomes more permeable due to the reduction of the pathway to evacuate the air.  

\begin{table}[H]
\centering
\resizebox{1\textwidth}{!}{\begin{tabular}{ccccccc}
	\hline
	Solids loading (wt.\%) & Porosity (\%) & Freezing Temperature ($^{\circ}$C) & Mean $d_{p}$ ($\mu m$) & $V_{Slurry}$ (ml) & Thickness (mm) & $k_{1}$ ($m^{2}$)
	\\ \hline \hline
	65 & 52 & -80 & 3.3 & 20 & 2.0 & 2.80\e{-14}\\
	50 & 68 & -80 & 7.3 & 20 & 2.0* & 2.96\e{-13}\\
	65 & 50 & -30 & 6.2 & 20 & 2.0* & 2.20\e{-13}\\
	65 & 50 & -80 &  3.3** & 18 & 1.6 & 2.40\e{-14}\\
	65 & 49 & -80 &  3.3** & 16 & 1.2 & 1.59\e{-14}\\
	\hline
\end{tabular}}
\caption{Summary of the main experimental conditions and the effects on porosity, mean pore size ($d_{p}$), and permeability ($m^{2}$).** mean pore size and * thickness of the tube not measured but considered equivalent to the sample frozen at -80$^{\circ}$C, 65\% solids loading, and 20 ml of volume of slurry ($V_{Slurry}$).}
\label{tab:StrucVSPermea} 
\end{table}

\begin{figure}[H]
	\centering
	\includegraphics[width=0.55\textwidth]{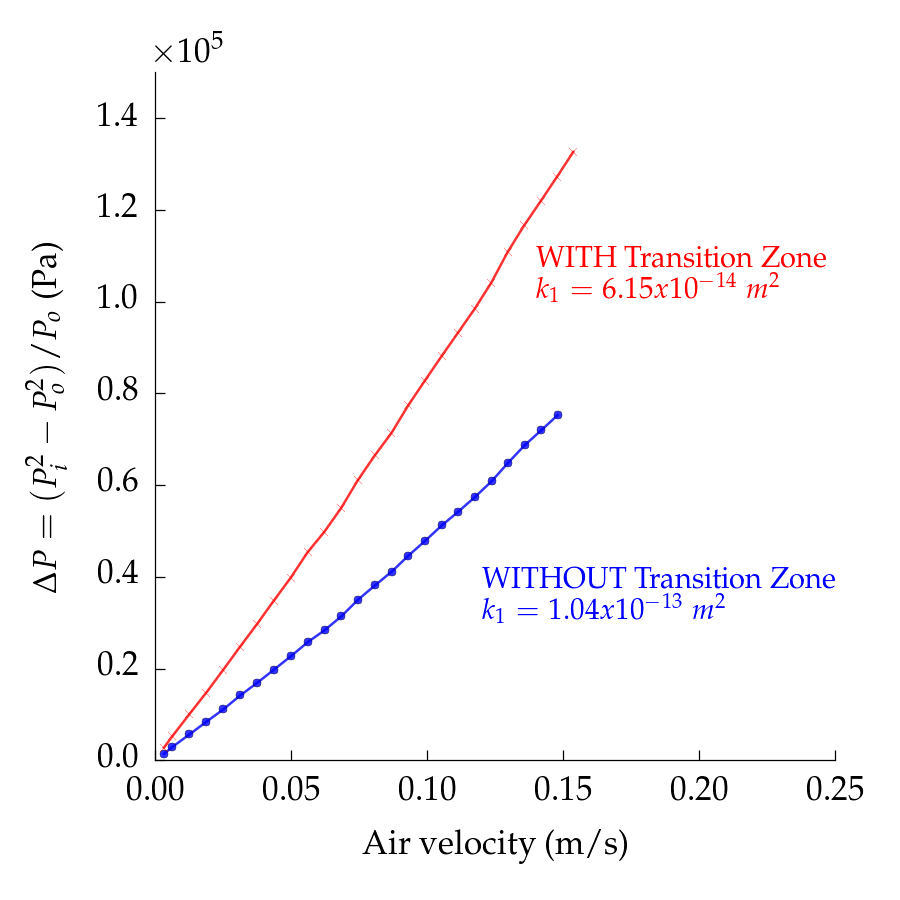}
	\caption{Effect of transition zone on pressure drop.}
	\label{fig:PressureDrop_TZ}
\end{figure}

Although, the external layer with finer and non-oriented porosity (referred here as transition zone) can facilitate the subsequent deposit of a dense coating \cite{Moon2003}, it might also act as a bottle-neck and consequently enhance the pressure drop. This effect can be observed in Fig. \ref{fig:PressureDrop_TZ} where a permeability test was performed in a tube before and after sanding off the external surface. The tube had a P(\%) of 45\%, a mean pore size of 3 $\mu m$, and an overall thickness around 2 mm. The specimen before the superficial treatment, and thus still with the transition zone, exhibited a higher pressure drop (i.e. lower $k_{1}$) compared with the sample after removing the superficial layer. The increase of permeability by a factor 6 highlights the role of this transition zone on permeability and draws the attention towards new hierarchical bilayared microstructures as the path to enhance the gas or liquid flow without an exaggerated increase in pressure drop.   

\section{Conclusions}

We developed a set-up to produce ice-templated tubes with accurate control of pore volume, pore size, and average thickness. Mechanical testing results showed a clear impact of pore volume and size on flexural strength, but results from radial crushing were inconclusive.  The permeability of tubes was successfully characterized, and we observed a high dependence on pore size.

The formation of a hierarchical microstructure obtained with this set-up could be beneficial in applications that require an asymmetric configuration such as solid oxide fuel cells, oxygen transport membranes, or water filtration. The external layer with finer and non-oriented porosity can facilitate the deposit of a dense coating and also increase the electrochemical reaction sites, while the unidirectional lamellar porosity minimizes the energetic losses reducing the effective transport pathway.

\section*{Acknowledgements}
The research leading to these results has received funding from  the CNRS and Saint-Gobain under BDI grant agreement 084877 for the Institut National de Chimie (INC), and from the European Research Council under the European Union's Seventh Framework Programme (FP7/2007-2013) / ERC grant agreement 278004 (project FreeCo).

\pagebreak

\bibliographystyle{model1-num-names}
\bibliography{library.bib}

\end{document}